# Supporting Instructors in Collaborating with Researchers using MOOClets

**Joseph Jay Williams**

Harvard University, HarvardX

125 Mt. Auburn St Rm. 422

Cambridge, MA 02138 USA

Joseph_jay_williams@harvard.edu

**Juho Kim**

MIT CSAIL

32 Vassar Street, 32-G707

Cambridge, MA, 02139

juhokim@mit.edu

**Brian C. Keegan**

Harvard University

430 Morgan Hall, Soldier's Field Rd.

Harvard, MA, 02138 USA

bkeegan@gmail.com

## Abstract

Most education and workplace learning takes place in classroom contexts far removed from laboratories or field sites with special arrangements for scientific research. But digital online resources provide a novel opportunity for large-scale efforts to bridge the real-world and laboratory settings which support data collection and randomized A/B experiments comparing different versions of content or interactions [2]. However, there are substantial technological and practical barriers in aligning instructors and researchers to use learning technologies like blended lessons/exercises & MOOCs as both a service for students and a realistic context to conduct research. This paper explains how the concept of a "MOOClet" can facilitate research-practitioner collaborations. MOOClets [3] are defined as modular components of a digital resource that can be implemented in technology to: (1) allow modification to create multiple versions, (2) allow experimental comparison and personalization of different versions, (3) reliably specify what data are collected. We suggest a framework in which instructors specify what kinds of changes to lessons, exercises, and emails they would be willing to adopt, and what data they will collect and make available. Researchers can then: (1) specify or design experiments that compare the effects of different versions on quantifiable outcomes. (2) Explore algorithms for maximizing particular outcomes by choosing alternative versions of a MOOClet based on the input variables available. We present a prototype survey tool for instructors intended to facilitate practitioner-researcher matches and successful collaborations.

**ACM Classification Keywords**
H.4 Information Systems Applications; H.5 Information interfaces and presentation; K.3.1 Computer Uses in Education; J.4 Social and Behavioral Sciences

**Introduction**
Whether in MOOCs or classrooms, many of the key affordances of digital resources have yet to be exploited in a way that simultaneously benefits instructors and researchers. This omission is striking since even the simple switch from an "offline" resource like pen-and-paper assignments and spoken lectures to online interactive exercises and webpages has the potential to change how practitioners and researchers in education interact. When implemented appropriately, detailed behavioral and user data can be collected from randomized experiments that compare the quantifiable educational benefits of different versions [3].

Digital resources can be further used to solicit feedback from diverse experts on concrete products (just by emailing a URL), and multiple versions rapidly created for comparison, personalization, and perpetual improvement. Researchers and teaching teams could mutually benefit from collaboration [4], via integration of research more directly into the development and modification of course materials.

**Barriers to Instructor-Researcher Collaboration**
Unfortunately, both instructors and researchers note that existing technology does not surmount many barriers to mutually beneficial collaboration. Several are listed below. These are synthesized from: A series of cross-disciplinary academic symposia and workshops; Discussions and structured interviews with MOOC instructional teams at Berkeley, Stanford, and Harvard; and our own work implementing research within the platforms EdX, Coursera, NovoEd, Khan Academy, Moodle, and ASSISTments.

1. Insufficient shared context to collaborate on aligning practical and scientific goals in concrete terms.
2. Difficulty in specifying and visualizing a vast set of potential modifications to educational materials.
3. Technical opacity in the affordances of platforms and development requirements for changes, exacerbated by risk averseness to bugs/glitches and high-pressure deadlines.
4. The challenge of scoping collaborative development rather than relinquishing editorial control over an entire course, or requiring substantial time to meetings.
4. Limited short-term payoffs for students in courses versus long-term benefit to research.
6. Difficulty understanding what data is collected and when it is available for analysis by scientists, and correspondingly instructors are uncertain about what personal participant data is shared and how to judge its ethicalness.

**Focusing collaboration around Experimental Comparisons within 'MOOClets'**
Drawing on our previously mentioned collaborative work, we explain the formal definition of a 'MOOClet'

[2] as a user-facing modular sub-component of an online or blended educational platform (e.g. a video, lesson, exercise, assignment, email, interactive study tool [1]) that can be authored, personalized, instrumented and experimentally varied, without requiring integration with underlying website architecture. The concept of a MOOClet is accompanied by a software design pattern for implementation across many platforms [2], a workflow for deciding details of a collaborative study, and a prototype interactive online rubric that instructors and researchers can use together to productively focus their discussion.

**Modularity.** Instead of discussions as the course level, discussions are targeted at potential modifications and data collection from modular components of educational experiences that are present in most courses regardless of goal and topic. Examples considered in past work [1] are exercises, quizzes, motivational videos, email reminders and suggestions, and lessons and interactive tools for practicing better study strategies.

**Flexible yet abstract technical implementation.** The design pattern for a MOOClet's definition and implementation in [2] provides guarantees to both researchers and instructors. Any component of a resource implemented as a MOOClet can have multiple versions created, and these can be experimentally compared yet dynamically adjusted by instructors to present preferred versions. The data collected in a MOOClet is pushed in real-time to a User Variable Store with differential privacy access, anonymity, but transparency about which variables are there and what their values are. Researchers therefore know what outcome variables and covariates are being collected, and can verify this information student-by-student as MOOClets are used.

We have shown [2] that most components of an online course or MOOC can be 'MOOCletized'. For example, any lesson, video, exercise, or email in EdX and Moodle. Researchers can conduct experimental comparisons of whatever different versions of a MOOClet they can creatively design, covering a broad range of potential levers for behavior, like adding or changing text, restructuring a lesson, including additional reflective questions [1].

**Align Experimentation & Personalization.** Moreover, MOOClets provide a mathematically formalization of how A/B experiments are equivalent to personalized delivery of different versions of a resource. The different versions of a MOOClet can be personalized based on any variables in the User Variable Store that come from outside sources of other MOOClets. This aligns a key behavioral science methodology of experimentation with work on intelligent tutoring systems and machine learning and artificial intelligence algorithms for adaptive presentation and recommendation of content, and application of classifiers or predictors to structured data. In addition, it aligns research efforts with a core instructional goal for digital technologies – providing individualized instruction. This approach has been taken in embedding experiments in K-12 math exercises on Khan Academy, videos on EdX, and teacher and medical professional development on NovoEd and Moodle [1].

**Experimental comparisons support practical evaluation of alternative designs *and* test**

**alternative hypotheses.** Researcher-practitioner alignment and collaboration is enabled by collaboratively designing experimental comparisons of different versions of MOOClets. These versions contain both practical design alternatives of instructional interest [3] *and* experimental contrasts designed to test alternative hypotheses about how people learn and interact with technology.

**Scoping modifiable course components.**
By focusing on just a few modular components of a course, instructors can be confident about will and won't be changed. They can know what data is being collected about students.

## Components of the Tool
The tool and examples of questions can be seen at http://tiny.cc/moocletcollaboration.

The questions in the tool are designed to elicit instructors and researchers respective beliefs about: (1) What quantifiable measures they want to improve, what kind of behaviors students engage in, and what data will be collected. (2) What modular components of the course instructors are interested in or opening to having modified and multiple versions experimented with or personalized, and so what experimental conditions researchers can design. E.g., homework exercises, text documents.

An example of a question included is "What components of the course would you like to improve? Or are you open to having modified?". All questions have both a free response open field – to get rich qualitative feedback and personal responses – as well as a list of options which is dynamically updated using responses of instructors.

To imagine one use case: Designers of an online course may care about a specific outcome, such as increasing retention. Findings from the psychology literature may point to appropriate metrics for measuring engagement as well as potential interventions to adopt [1]. The set of interventions might include additional advice videos from the instructor on the landing page or dashboard, messages embedded above exercises, or email reminders [3]. Measures of engagement might include self-reported mind-wandering, time spent on a page, and number of attempts made on a difficult problem.

## References

[1] Williams, J. J., Maldonado, S., Williams, B. A., Rutherford-Quach, S., & Heffernan, N. (2015). How can digital online educational resources be used to bridge experimental research and practical applications? Embedding In Vivo Experiments in "MOOClets". *Paper to be presented at the Spring 2015 Conference of the Society for Research on Educational Effectiveness, Washington, D. C.*

[2] Williams, J. J., Li, N, Kim, J., Whitehill, J., Maldonado, S., Pechenizkiy, M., Chu, L., & Heffernan, N. (under review). *The MOOClet Framework: Improving Online Education through Experimentation and Personalization of Modules*. Tiny.cc/moocletframework

[3] Williams, J.J. & Williams, B. A. (2013). Using Randomized Experiments as a Methodological and Conceptual Tool for improving the Design of Online Learning Environments. *Paper presented at the Data Driven Education Workshop at the Conference on Neural Information Processing Systems.*

Wuchty, S., Jones, B. F., & Uzzi, B. (2007). The increasing dominance of teams in production of knowledge. *Science, 316*(5827), 1036-1039.